%% file: paperID.tex
\documentclass[runningheads]{llncs}

\usepackage{graphicx}
\usepackage{amssymb}
\usepackage{amsmath,tabularx}
\usepackage{xcolor}
\usepackage{multirow}
\usepackage{siunitx}
\usepackage{booktabs}
\usepackage{etoolbox}
\usepackage{color}
\usepackage{lipsum}
\usepackage{tabu}
\usepackage{dsfont}
\usepackage{subcaption,siunitx,booktabs}

\usepackage[symbol]{footmisc}

\definecolor{todo}{rgb}{1.0, 0., 0.}

\newcommand\asteriskfill{\leavevmode\xleaders\hbox{$\ast\ $}\hfill\kern0pt}
\newcommand{\comment}[1]{}
\begin{document}

\robustify\bfseries
        
\title{Longitudinal Self-Supervision for \\ COVID-19 Pathology Quantification}

\titlerunning{Longitudinal Self-Supervision for COVID-19}

\author{Tobias Czempiel\inst{1}, Coco Rogers\inst{1,2}, Matthias Keicher\inst{1}, Magdalini Paschali\inst{3}, Rickmer Braren\inst{4}, Egon Burian\inst{4}, Marcus Makowski\inst{4}, Nassir Navab\inst{1,6}, Thomas Wendler\inst{1}, Seong Tae Kim\inst{5,}\thanks{\textit{Corresponding author (st.kim@khu.ac.kr)}}}

\authorrunning{Czempiel et al.}

\institute{
Computer Aided Medical Procedures, TUM, M{\"u}nchen, Germany
\and
Department of Computer Science, LMU, M{\"u}nchen, Germany
\and
Department of Psychiatry and Behavioral Sciences, Stanford, Stanford, USA
\and
Department of Diagnostic and Interventional Radiology, TUM, Germany
\and
Department of Computer Science and Engineering, Kyung Hee University, Korea
\and
Computer Aided Medical Procedures, Johns Hopkins University, Baltimore, USA 
}

\maketitle              
\begin{abstract}
Quantifying COVID-19 infection over time is an important task to manage the hospitalization of patients during a global pandemic. Recently, deep learning-based approaches have been proposed to help radiologists automatically quantify COVID-19 pathologies on longitudinal CT scans. However, the learning process of deep learning methods demands extensive training data to learn the complex characteristics of infected regions over longitudinal scans. It is challenging to collect a large-scale dataset, especially for longitudinal training. In this study, we want to address this problem by proposing a new self-supervised learning method to effectively train longitudinal networks for the quantification of COVID-19 infections. For this purpose, longitudinal self-supervision schemes are explored on clinical longitudinal COVID-19 CT scans. Experimental results show that the proposed method is effective, helping the model better exploit the semantics of longitudinal data and improve two COVID-19 quantification tasks.

\keywords{COVID-19 \and Longitudinal Analysis \and Self-Supervision}
\end{abstract}

\input{1.Introduction}
\input{2.Methodology}
\input{3.ExperimentalSetup}
\input{4.Results}
\input{5.Conclusion}

\bibliographystyle{ieeetr}
\bibliography{bibfile}

\end{document}

%% file: 1.Introduction.tex
\section{Introduction} 
COVID-19 has caused a global pandemic with more than 437 million cases and more than 5.96 million deaths worldwide (as of March 1st, 2022)\footnote{https://coronavirus.jhu.edu/map.html}. The infection can have very mild symptoms, but in some cases, it develops into a severe condition that requires intensive medical treatment \cite{Yu2020}. Chest computed tomography (CT) enables quantifying the infection burden of the lungs by showing specific pathological patterns such as consolidation and ground-glass opacity (GGO). Thus, CT can be used to identify patients at high risk of a severe course requiring admission to the ICU \cite{Burian2020}. 
Although segmentation of infected regions is essential \cite{Zhou2020}, manual CT volume segmentation is very time consuming and limited by high inter-observer and intra-observer variability \cite{Wang2020}.

Several deep learning-based studies have been presented to automatically segment and analyze COVID-19-CT scans \cite{Zhou2020,Wang2020,Fan2020,Shan2021}. The infections vary between patients' size, position, and shape. In addition, the infected areas have a variety of textures and subtle anatomical boundaries \cite{Wang2020,Fan2020}.

To assess disease progression and response to treatment, the analysis of longitudinal CT scans from the patient is essential. Several medical studies have investigated the use of longitudinal data to improve segmentation performance in \cite{Birenbaum2017,denner2020spatiotemporal}. Kim et al. \cite{kim2021longitudinal} introduced a longitudinal model to analyze COVID-19 progression over time and monitor the recovery process and response to various medical treatments. However, deep learning approaches require a large amount of labeled data to learn the complex texture and boundaries of infections. For the longitudinal data collection, this problem is even more challenging. 

Self-supervised learning schemes have been widely explored to improve the results in low data regimes. In self-supervised learning, the central methodological choice revolves around the design of an adequate pretext task for supervision from the data  \cite{Pathak_2016_CVPR,context_restoration,model_genesis,ZHAO2021102051}. Most self-supervised learning approaches have been studied only for static data. Only a few research efforts have been devoted to devising self-supervised learning on longitudinal data \cite{ZHAO2021102051}. However, previous works mainly focus on improving the classification task.

In this paper, we develop self-supervised pretext tasks for the longitudinal domain to improve the results of longitudinal deep neural networks under a limited amount of data. The proposed self-supervised pretext task helps the model to better understand the CT scans' longitudinal characteristics for both segmentation and classification. Our main contributions are:

\begin{itemize}

\item To the best of our knowledge, this is the first study to explore self-supervised learning for the analysis of longitudinal COVID-19 CT scans. By exploiting pretext tasks for pretraining, the longitudinal model learns to use the data more effectively, resulting in performance improvements in two downstream tasks for segmentation and classification.
\item We present a comprehensive analysis to verify our novel training scheme as a pioneer work in self-supervised learning for longitudinal COVID-19 pathology quantification. We evaluate our longitudinal model, trained on data from the first COVID-19 wave on an independent test set acquired during the second wave.

\end{itemize}

%% file: 2.Methodology.tex
\section{Longitudinal Self-Supervised Learning}
Longitudinal segmentation, which uses two concatenated scans as input to exploit both spatial and temporal cues (spatio-temporal) cues, has shown very promising results for COVID-19 analysis \cite{kim2021longitudinal,foo2021interactive}. The major limitation of longitudinal segmentation is the limited availability of annotated training data to sufficiently learn the underlying distribution, resulting in performance and generalization drops.
In this section, we propose a new approach to overcome these limitations by introducing a longitudinal self-supervised learning scheme for COVID-19 analysis. As shown in Fig. 1, our framework consists of two components: 1) a longitudinal self-supervised pretraining and 2) a longitudinal analysis for downstream tasks (i.e., segmentation and classification). 

\begin{figure}[t]
\centering
\includegraphics[width=\textwidth]{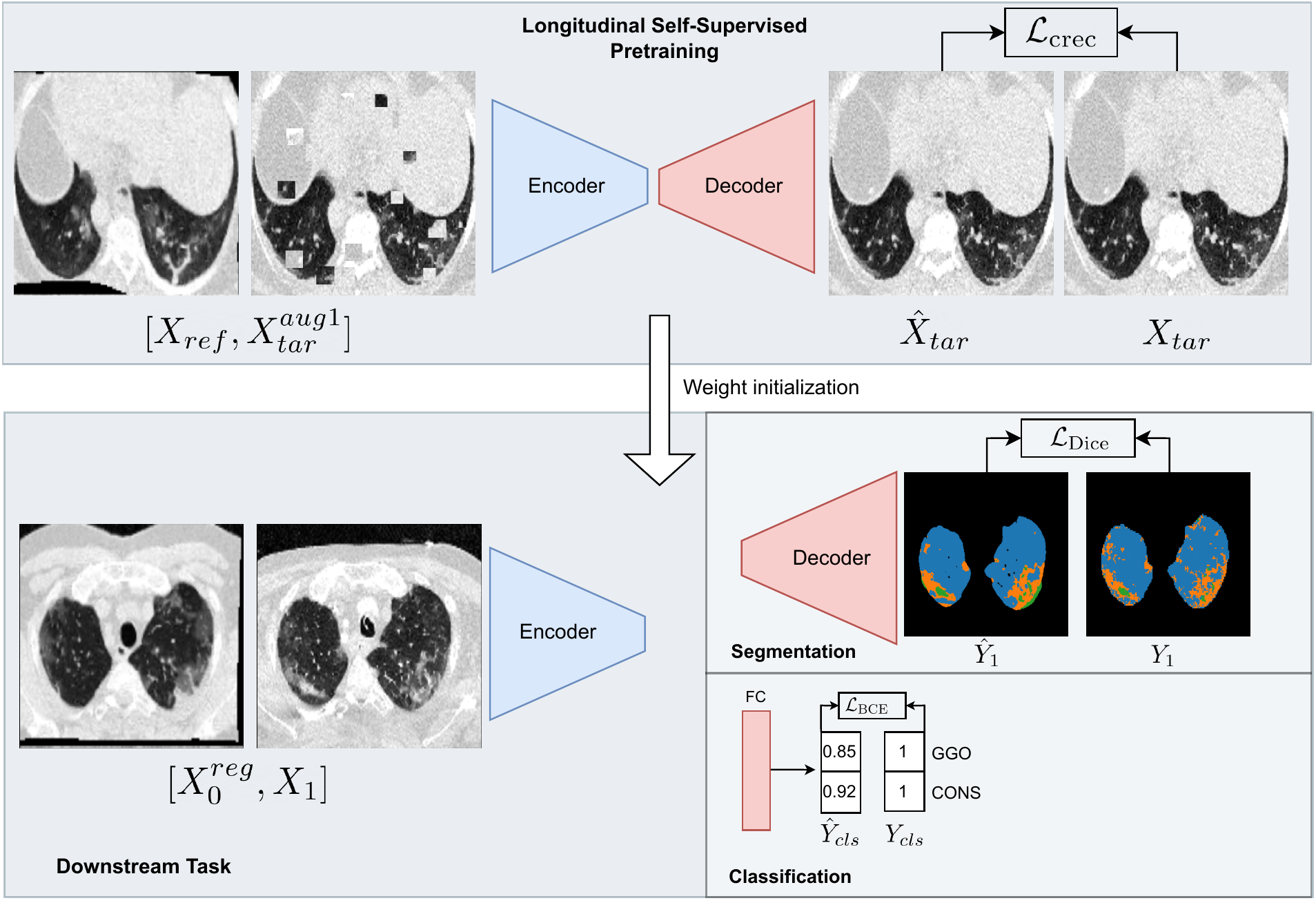}
\caption{The overall framework of the proposed method including longitudinal self-supervised learning and two downstream tasks.}
\label{fig:reconstruction}
\end{figure}

\subsection{Longitudinal Self-Supervision}
\label{self-supervised_pretraining}
Through self-supervised pretraining, the model can learn descriptive semantic features that can help improving different downstream tasks \cite{context_restoration}. We, therefore, propose a new longitudinal self-supervised pretraining model based on the image restoration task.
In this study, a baseline longitudinal network \cite{kim2021longitudinal} is defined based on a fully convolutional DenseNet \cite{densenet} consisting of a downsampling path with five transition down blocks and an upsampling path with five transition up blocks, respectively.

The input for the model is the concatenation of two registered 2D CT slices from two different time points (i.e., reference scan and follow-up scan). 
Before we feed the slices in our method, we perform a 3D registration of the reference scan to the follow-up scan [$X_{ref}$, $X_{tar}$] to align the corresponding slices of the volumes with each other.
For the pretext task the target scan is modified through data augmentation [$X_{ref}$, $X_{tar}^{aug}$]. The objective of the pretext model is the restoration of the target CT scan $\hat X_{tar}$. The network, therefore, does not require any additional annotation as ground truth supervision. 

To train our method we choose a combination of the Mean Squared Error (MSE) loss between the restored and the original CT slice and a patch loss focusing on the reconstruction results purely on the augmented image regions. A mask $M$ of the modified patches is used to calculate the patch loss which is  weighted equally to the MSE loss. The resulting loss function is defined as 

\begin{equation}
\mathcal{L}_{\text{crec}} = 0.5 * \mathcal{L}_{\text{MSE}} + 0.5 * \mathcal{L}_{\text{Patch}}, 
\end{equation} 
with $\mathcal{L}_{\text{MSE}} = \mathcal{L}_{\text{MSE}}(\hat X_{tar}, X_{tar})$ and $
\mathcal{L}_{\text{Patch}} = \mathcal{L}_{\text{MSE}}(\hat X_{tar} * M, X_{tar} * M)$. This loss is designed to encourage the model to focus on correctly reconstructing the modified patches. 

\begin{figure}[t]
\centering
\includegraphics[width=\textwidth]{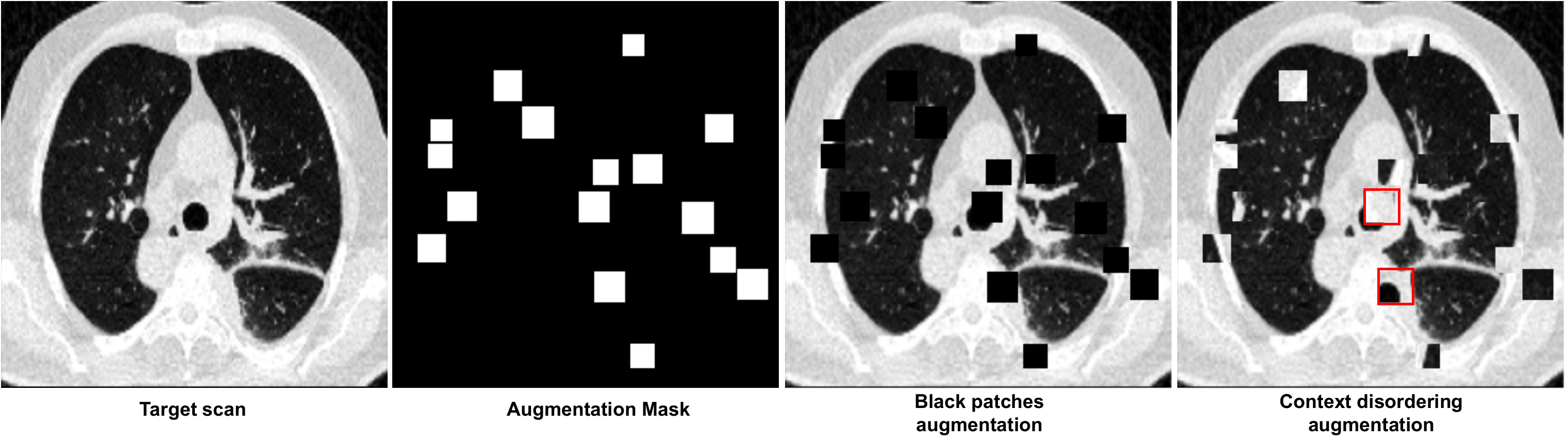}
\caption{Example of Black patches and Context disordering data augmentation. From left to right starting with the non-augmented target scan, the augmentation mask $M$, the target scan augmented with Black patches and the target scan augmented with Context disordering where the red boxes indicate an example of two swapped patches.}
\label{fig:augmentations}
\end{figure}

\subsection{Data Augmentations}
\label{data_augmentations}

For the self-supervised pretext task, two different well known augmentations tasks \cite{context_restoration,Pathak_2016_CVPR} are adopted to the longitudinal task: Context disordering and Black patches. The idea is to pretrain the model on data from the same domain and thereby learn useful semantic information. In addition, the longitudinal nature of the data is used and the model has the chance to use the information present in the reference scan to restore the augmented follow up scan. As a result, the model learns to use the information from the reference scan to improve the restoration the task. 
Both augmentations are based on modified square patches positioned randomly, ensuring that none of them overlap. The size of the patches as well as the number of augmentations per image is randomly selected for each datapoint.

\subsubsection{Context disordering.}
For Context disordering \cite{context_restoration}, two patches in $X_{tar}$ are randomly selected and their positions are swapped. This operation is repeated several times and changes the spatial information while preserving the intensity distribution. An example of the Context disordering transformation is shown in Figure \ref{fig:augmentations}. The result of this process is the input pair [$X_{ref}$, $X_{tar}^{aug1}$].
To complete the task correctly the model has to learn to identify the changed patches and reverse the augmentation of the context correctly. 

\subsubsection{Black patches.} 
In the black patches augmentation \cite{Pathak_2016_CVPR} we place black patches at random locations on the target scan. This process results in the input pair [$X_{ref}$, $X_{tar}^{aug2}$] visualized in Figure \ref{fig:augmentations}.
To complete the task correctly the model has to reconstruct the original image from the augmented image based on the surrounding context of a augmented region and longitudinal data.

%% file: 3.ExperimentalSetup.tex
\section{Experimental Setup}

\subsection{Datasets}

The dataset used for training and validation is an in-house dataset containing longitudinal low-dose CT volumes from 37 patients with a positive PCR from the first COVID-19 wave (March-June 2020). 
The CTs were separated 17 $\pm$ 10 days (1-43 days) and were taken at admission and during the hospital stay (33 $\pm$ 21 days, 0-71 days).  
All scans were performed using two different CT devices (IQon Spectral CT and iCT 256, Philips, Hamburg, Germany) with the same parameters (X-ray current 140-210 mA, voltage 120 kV peak, slice thickness 0.9mm, no contrast media) and covered the complete lung. The data collection was performed with the approval of the TUM institutional review board (ethics approval 111/20 S-KH). 
The dataset was annotated at a voxel-level by a single expert radiologist (5 years experience), generating
pathology masks with three different classes: healthy lung (HL), ground-glass opacity (GGO) and consolidation (CONS).
The dataset was split into training and validation set with an 80/20 split resulting in 30 patients with 66 volumes in the training and 7 patients with 20 volumes in the validation set. 

For testing, we collected an independent dataset acquired during the second COVID-19 wave (June-November 2020) with the same CT Scanners. It contains 50 longitudinal CTs and labeled pathology masks from 25 patients (i.e., 25 follow-up volumes which consist of 3,273 slices in total).
The CTs were separated 18.56 $\pm$ 10.7 days (4-50 days) and were taken during the hospital stay (36 $\pm$ 18 days). All of the 25 patients needed intensive care and 7 patients died. 

For preprocessing the following steps are performed. We first cropped the CT scans to the lung region and further applied a min-max normalization before using them in our models.
To align CTs at different time points, we applied a BSpline Transformation based deformable registration algorithm \cite{registration}. The BSpline Transformation is defined using a sparse set of grid points that are placed over the fixed domain of the image.
The reference scan lung mask $M_{0}$ is registered to the lung mask of the target scan $M_{1}$. The resulting transformation function is defined as $R_{M_{0} \to M_{1}}(\cdot)$. Based on this function, the transformations are applied to the respective CT scans as $X_{0}^{reg} = R_{M_{0} \to M_{1}}(X_{0})$. In this process, only the reference volume $X_{0}$ is modified while leaving the target volume untouched. 

\subsection{Longitudinal COVID-19 Pathology Quantification}
As downstream tasks, we explore a medical segmentation and classification task for longitudinal COVID-19 pathology quantification. 

\subsubsection{Longitudinal segmentation.}
For the segmentation task, we follow Kim et al. \cite{kim2021longitudinal} as a baseline model, an early fusion network that using as input slices of two concatenated registered scans of a patient from two different time points.
This allows the network to capture temporal changes between the two scans. The reference scan (t=0) is always registered with the follow-up scan (t=1) [$X_{0}^{reg}$, $X_{1}$]. The pathologies are then segmented on the follow-up scan $X_{1}$. For patients with more than two scans available, we registered all possible combinations from the past to the future in order to retain the true temporal progression of the disease.
Our model has four output channels for the classes: background, healthy lung, GGO, and consolidation.

\subsubsection{Longitudinal classification.}
For the classification task, the goal of the model is to predict the presence of GGO and consolidation from the longitudinal pair of input slices for the follow up scan in a slice-level. We replace the original decoder with a single linear layer using the bottleneck features of the encoder as the input and the class-level prediction of pathological presence as the output. 

\subsection{Model Training}
\subsubsection{Longitudinal self-supervised learning.}
We run the longitudinal self-supervised pretraining for 100 epochs, using the validation loss as the early stopping metric with a patience of 5 epochs. The model with the lowest validation loss is selected for testing. The batch size is set to 4 and the Adam optimizer \cite{adam2014} is used with a learning rate of 0.0001.
We randomly select the amount of modified patches in an interval [16,25] on an instance level.
A longitudinal and a static version of each pretext task is trained for comparison.

\subsubsection{Longitudinal segmentation.} \label{segmentation}
For the training of th longitudinal segmentation we initialize the model with the weights from the longitudinal self-supervised pretraining task.
In this way, we want to transfer the knowledge learned in the pretext tasks to our segmentation model.
The model is trained over 30 epochs using the Adam optimizer \cite{adam2014} with a learning rate of 0.0001 and a Dice loss for segmentation. The model with the highest mean Dice score in the validation set is selected. The static segmentation runs are initialized by the weights from the static self-supervised learning. For the evaluation of the segmentation, the different segmentation models are tested on the independent test dataset.
\subsubsection{Longitudinal classification}
Similarly to the segmentation \ref{segmentation}, we initialize our encoder with the weights from the pretraining task. We train our model with the Binary Cross-Entropy-Loss for the pathology presence task to account for the multi label nature of our model. In that way, the presence of the GGO and Consolidation is predicted independently from each other.

%% file: 4.Results.tex
\section{Results and Discussion} \label{results}

\begin{table}[t]
\centering
\caption{Dice scores for the different classes of static and longitudinal segmentation without pretraining, with Black patches pretraining, and with Context disordering pretraining.}
\label{tab:long_pretraining}
\resizebox{\textwidth}{!}{%
\begin{tabular} {@{}l@{\hskip 0.25in}c@{\hskip 0.25in}c@{\hskip 0.25in}c@{\hskip 0.25in}c@{}}
\toprule
\textbf{Segmentation}& \textbf{Healthy} & \textbf{GGO} & \textbf{CONS} & \textbf{Overall} \\ 
\midrule \midrule \textbf{Static} &  & \\ 
\begin{tabular}[c]{@{}l@{}}No pretraining\end{tabular} & 0.658 $\pm$ 0.039 & 0.515 $\pm$ 0.037 &  0.348 $\pm$ 0.037 & 0.623 $\pm$ 0.039  \\  \midrule
\begin{tabular}[c]{@{}l@{}}Black patches\end{tabular} &    0.674 $\pm$ 0.038 & \textbf{0.524 $\pm$ 0.036} &  0.357 $\pm$ 0.037 &  0.635 $\pm$ 0.039\\ \midrule
\begin{tabular}[c]{@{}l@{}}Context disordering\end{tabular} & \textbf{0.674 $\pm$ 0.038} & 0.519 $\pm$ 0.034 &  \textbf{0.364 $\pm$ 0.033}  & \textbf{0.636 $\pm$ 0.039} \\
\midrule
\midrule
\textbf{Longitudinal} &  & \\
\begin{tabular}[c]{@{}l@{}}No pretraining\end{tabular} & 0.675 $\pm$ 0.041 & 0.531 $\pm$ 0.034 & 0.372 $\pm$ 0.034 & 0.644 $\pm$ 0.039 \\ \midrule
\begin{tabular}[c]{@{}l@{}}Black patches\end{tabular} &  0.683 $\pm$ 0.038 & 0.530 $\pm$ 0.035 & 0.398 $\pm$ 0.038 & 0.652 $\pm$ 0.038 \\ \midrule
\begin{tabular}[c]{@{}l@{}}Context disordering\end{tabular} & \textbf{0.684 $\pm$ 0.040} &  \textbf{0.541 $\pm$ 0.034} & \textbf{0.413 $\pm$ 0.036} & \textbf{0.659 $\pm$ 0.037} \\
\bottomrule
\end{tabular}%
}
\end{table}

\begin{table}[t]
\centering
\caption{Classification results with and without Context disordering pretraining in the longitudinal network.}
\label{classificationtable}
\resizebox{\textwidth}{!}{%
\begin{tabular}{@{}l@{\hskip 0.15in}c@{\hskip 0.15in}c@{\hskip 0.15in}c@{\hskip 0.15in}c@{}}
\toprule
\textbf{\begin{tabular}[c]{@{}l@{}}Classification\end{tabular}} & \textbf{AUC} & \textbf{\begin{tabular}[c]{@{}l@{}}Accuracy\\
(Overall)
\end{tabular}} & \textbf{\begin{tabular}[c]{@{}l@{}}Accuracy\\ (GGO)\end{tabular}}   & \textbf{\begin{tabular}[c]{@{}l@{}}Accuracy\\ (CONS)\end{tabular}}  \\ \midrule \midrule
\begin{tabular}[c]{@{}l@{}}No pretraining\end{tabular} & 0.817  $\pm$ 0.032 & 0.705 $\pm$ 0.023 & 0.670 $\pm$
0.047 & 0.740 $\pm$
0.028  \\ \midrule
\begin{tabular}[c]{@{}l@{}}Context disordering\end{tabular}     & \textbf{0.844 $\pm$ 0.023}      & \textbf{0.755  $\pm$ 0.021} & \textbf{0.747 $\pm$ 0.056} & \textbf{0.763  $\pm$ 0.033}  \\
\bottomrule
\end{tabular}
}
\end{table}

In Table~\ref{tab:long_pretraining} we compared the segmentation results of Covid-19 pathologies between a static and longitudinal architecture and with and without pretraining tasks. For the static model, we see that compared to no pretraining (62.3\%) both the black patches (63.5\%) and the context disordering pretraining (63.6\%) have a positive effect on the segmentation results overall. The longitudinal model without pretraing improves the results of all static models over each class and overall (64.4\%) by 1\% and compared to the static model without pretraining by 2\%. 
Further, we observe that the Black patches longitudinal pretraining model improves the results over the longitudinal model further on all classes. The context disordering longitudinal model performs best and we see an increase in the important pathology classes of 1\% for GGO and 3\% for CONS. Also the context disordering longitudinal model achieves the best overall results (65.9\%).

For the classification results in Table~\ref{classificationtable} we observe again that the Context disordering improves the results of the longitudinal model with a margin of 2.7\% in AUC and 5\% in Accuracy compared to the longitudinal model without pretraining. This implies that the proposed self-supervised learning helps the model to better exploit the semantics of longitudinal data.

\begin{figure}[t]
\centering
\includegraphics[width=\textwidth]{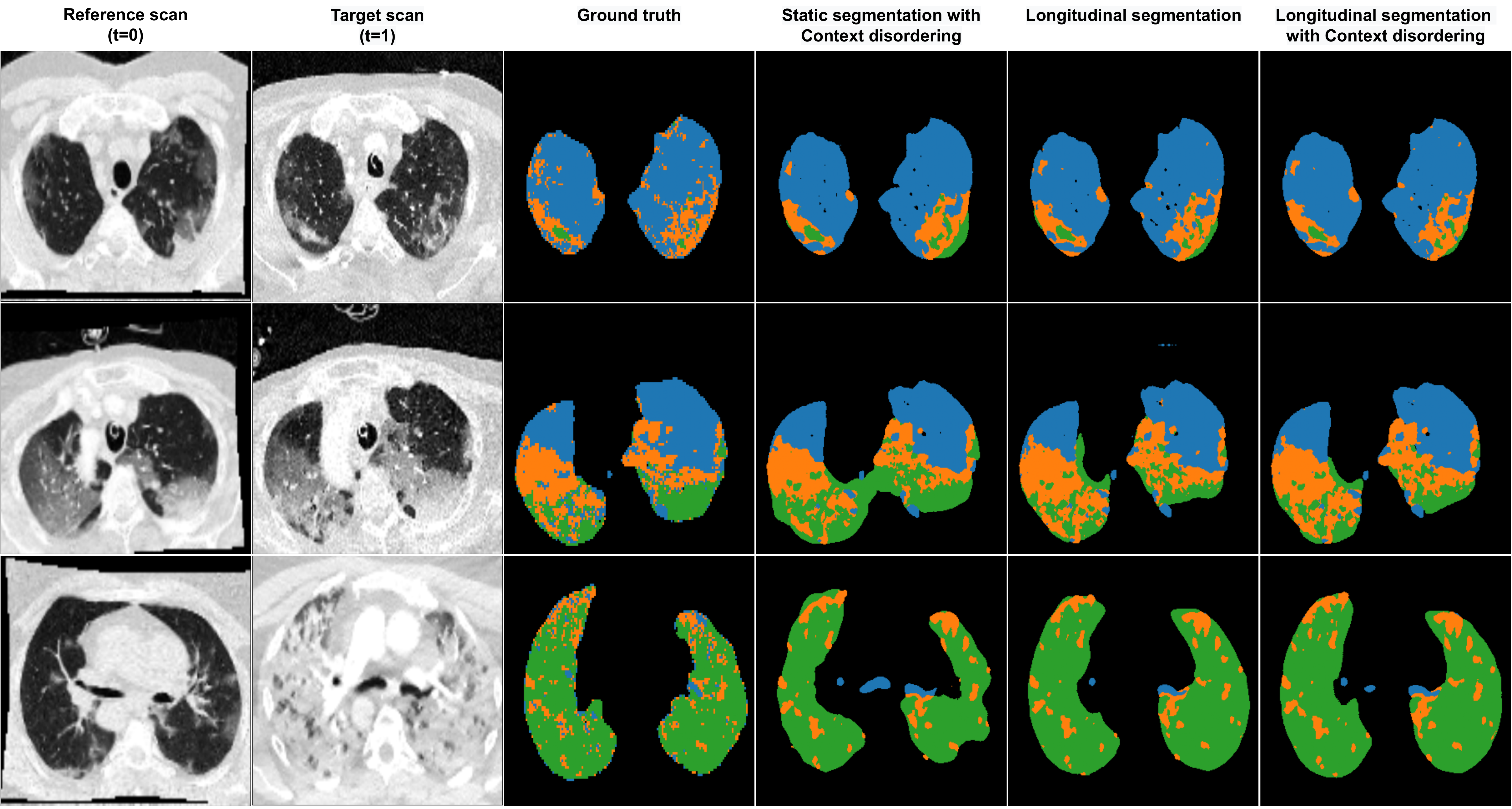}
\caption{Qualitative results from three different patients for the static segmentation pretrained with Context disordering, the longitudinal segmentation without pretraining and the longitudinal segmentation pretrained with Context disordering. Always the target scan is segmented. In the segmentation masks, blue, orange, and green indicate healthy lung, GGO, and consolidation, respectively. For the longitudinal network, both scans (reference and target) are used as input, for the static segmentation only the target scan.}
\label{fig:qualitative_results}
\end{figure}

%% file: 5.Conclusion.tex
\section{Conclusion}
In this study, we proposed a new self-supervised learning scheme to address the difficulties to collecting a large size of longitudinal COVID-19 data. By pretraining the longitudinal model using the proposed pretext tasks, we show that the model can better understand and exploit the nature of longitudinal scans for two downstream tasks. Experiments were conducted for evaluating the model, which was trained on the data collected in the first COVID-19 wave, on independent test data in second COVID-19 wave. The results verified the effectiveness of the proposed longitudinal self-supervision in COVID-19 pathology quantification. We will further investigate the clinical usability of the method by extending it to other longitudinal medical analysis tasks.

\section*{Acknowledgements}
\label{sec:acknowledgements}
\noindent
This paper was funded by the Bavarian Research Foundation (BFS) under grant agreement AZ-1429-20C. S.T. Kim was partially supported by National Research Foundation of Korea (NRF) grant funded by the Korea Government (MSIT) under Grant 2021R1G1A1094990. 